\documentclass[amsmath, amssymb,twocolumn,superscriptaddress,showpacs,showkeys,prb]{revtex4-1}
\usepackage{graphicx}

\begin{document}
\title{Influence of two-level fluctuators on adiabatic passage techniques}
\author{Nicolas \surname{Vogt}}
\affiliation{Institut f\"ur Theorie der Kondensierten Materie,  \\ Karlsruhe Institute of Technology, D-76128 Karlsruhe, Germany}
\affiliation{Institut f\"ur Theoretische Festk\"orperphysik, \\ Karlsruhe Institute of Technology, D-76128 Karlsruhe, Germany}
\affiliation{DFG-Center for Functional Nanostructures (CFN), Karlsruhe Institute of Technology, D-76128 Karlsruhe, Germany}
\author{Jared H. \surname{Cole}}
\affiliation{Chemical and Quantum Physics, School of Applied Sciences, \\ RMIT University, Melbourne, Australia}
\affiliation{Institut f\"ur Theoretische Festk\"orperphysik, \\ Karlsruhe Institute of Technology, D-76128 Karlsruhe, Germany}
\author{Michael \surname{Marthaler}}
\affiliation{Institut f\"ur Theoretische Festk\"orperphysik, \\ Karlsruhe Institute of Technology, D-76128 Karlsruhe, Germany}
\affiliation{DFG-Center for Functional Nanostructures (CFN), Karlsruhe Institute of Technology, D-76128 Karlsruhe, Germany}
\author{Gerd \surname{Sch\"on}}
\affiliation{Institut f\"ur Theoretische Festk\"orperphysik, \\ Karlsruhe Institute of Technology, D-76128 Karlsruhe, Germany}
\affiliation{DFG-Center for Functional Nanostructures (CFN), Karlsruhe Institute of Technology, D-76128 Karlsruhe, Germany}
\begin{abstract}
We study the process of Stimulated Raman Adiabatic Passage (STIRAP) under the influence  of a non-trivial solid-state environment, particularly the effect of two-level fluctuators (TLFs) as they are frequently present in solid-state 
devices. 
When the amplitudes of the driving-pulses used in STIRAP are in resonance with the level spacing of the fluctuators the quality of the protocol, i.e., the transferred population decreases sharply. 
In general the effect can not be reduced by speeding up the STIRAP process. 
We also discuss the effect of a structured noise environment on the process of Coherent Tunneling by Adiabatic Passage (CTAP). 
The effect of a weakly structured environment or TLFs with short coherence times on STIRAP and CTAP can be described by the Bloch-Redfield theory. For a strongly structured environment a higher-dimensional approach must be used, where the TLFs are treated as part of the system.
\end{abstract}
\pacs{74.50.+r, 03.65.Yz, 32.80.Qk, 03.67.Lx}
\keywords{superconducting three-state system, STIRAP, decoherence, Josephson junctions, two-level fluctuators}

\date{February 2, 2012}
\maketitle
\section{Introduction}
Much of the recent excitement in low-temperature device research, for example the growing field of \emph{circuit quantum electrodynamics}~\cite{Astafiev2007, Wallraf2008, Martinis2009, Gross2010,Marthaler2011}, centers around mapping quantum optics concepts to solid-state systems. 
An example to be discussed in the present work is Stimulated Raman Adiabatic Passage (STIRAP), a quantum control scheme to transfer population between two levels in an adiabatic rotation via a third level.
It has been widely studied in the field of quantum optics, its  advantage lying in the robustness against external perturbations and imperfections in the control pulses \cite{Bergmann1998}. 
STIRAP, therefore, has also been proposed for achieving single-qubit rotations in quantum information experiments \cite{Kis2002}. 
Specific protocols were discussed for superconducting qubits, charge qubits~\cite{ManganoSiewert2008}, phase qubits~\cite{Paspalakis2004a} and the quantronium\cite{Siewert2010, Falci2011}. Experiments on electromagnetically induced transparency and the formation of a dark state, both effects closely related to STIRAP, have also been performed ~\cite{Hakonen2009, Astafiev2010, Wilson2011}.
A related scheme called Coherent Tunnelling by Adiabatic Passage (CTAP) exists for three neighbouring quantum dots\cite{Greentree2004} where the tunnel matrix elements themselves are modulated to achieve an adiabatic transition.

The description of decoherence, both in quantum optics and circuit QED systems, is frequently based on a master equation approach as described by the Lindblad form. This formulation is sufficient for the structure-less environment which the vacuum provides in quantum optics, however, the solid-state environment can be much more complicated. In addition to the effects of the electric circuit, typically dominated by an Ohmic response, experiments displayed strong signatures of low-frequency or $1/f$ noise\cite{Clarke2008}.  This background noise may be described by a bath of independent two-level-fluctuators (TLFs)\cite{Shnirman2005, Schriefl2006}, however, recent work on qubits 
demonstrated the coupling to individual TLFs\cite{Martinis2005}. Qubits have be used to
 detect and manipulate them\cite{ Lupa2009,  Cole2010, *LisenfeldMuller2010}, yet their true microscopic nature is still unknown\cite{Cole2010}. The presence of TLFs in a solid-state environment  requires more complex models than are possible with the conventional Lindblad treatment.

In this paper we study how decoherence effects from a structured environment influence the efficiency,
i.e., the transferred population, in the STIRAP process.
When analysing these problems we compare different theoretical approaches.
We start with a short review what is known about the effect of noise 
on the STIRAP protocol as long as it is treated by a master equation with Lindblad damping terms. 
Next we introduce the model with TLFs and show how a weakly structured noise environment 
can be treated by the Bloch-Redfield equation for the three-level system. 
For a strongly structured noise environment, e.g., due to TLFs with long 
life times, the analysis of a higher-dimensional system formed by the three-level system 
and the TLFs is needed. 

We arrive at the following conclusions: 
As long as the dissipation arises due to wide-band noise its effect can be reduced by
using strong driving pulses. The situation changes for a structured bath, e.g., 
due to the presence of TLFs. If TLFs couple to the three-level system  the transferred population decreases sharply when the amplitude of the driving-pulses used in STIRAP coincides with the level spacing of the fluctuator. In general this effect can not be reduced by speeding up the STIRAP process. For an efficient STIRAP protocol the driving pulse amplitude 
must be chosen sufficiently low or, as long as the rotating wave approximation remains valid, high compared to the TLF frequency.
In a final section we study how the transferred population in a CTAP process is affected by a TLF coupling to the barrier-tunnelling amplitudes.

\section{STIRAP with Lindblad damping}
In this section we discuss the effects of noise on the STIRAP process in the frame of a master equation with Lindblad damping terms. As long as this description is valid, it turns out to be
advantageous to use strong driving pulses for the STIRAP protocol. In subsequent sections we will generalize the analysis to the case where the environmental bath is structured and the Lindblad theory is no longer sufficient.

\begin{figure}[th!]
	\vspace{0.5cm}
	\includegraphics[scale=0.4]{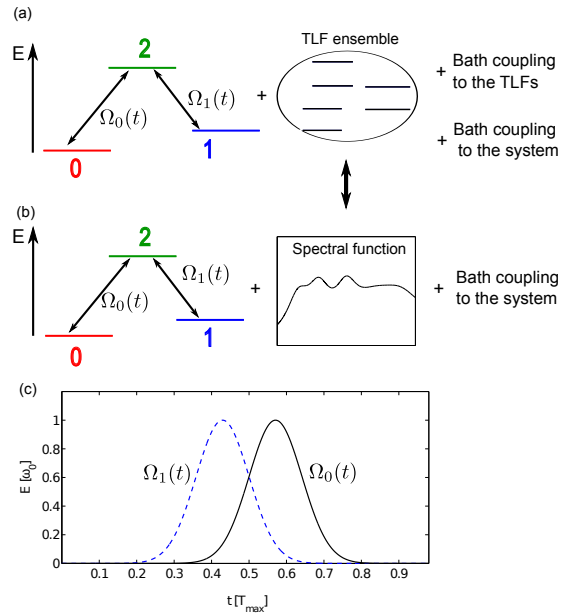}
	\caption{ \label{fig:TLFensemble} (a) and (b): The STIRAP process involves steering the population within a three-state system (states 0, 1 and 2) using applied laser (or microwave) driving fields with a time dependent Rabi frequency $\Omega_0(t)$ and $\Omega_1(t)$.  In a solid state environment, the three-level system can couple to non-trivial environmental modes stemming from two-level solid-state defects.  This effect can be modelled in two alternative ways (illustrated conceptually here, via explicitly treating the coherent defects in the environment (a) or via an effective spectral model using Bloch-Redfield theory (b).  While the former is `exact', it is computationally expensive.  The later scales better but is only applicable in certain parameter regimes. \\	
	(c): In order to realize the STIRAP protocol, the effective Rabi frequency of the driving pulses must be varied as a function of time,  plotted here for the counter-intuitive pulse sequence.}
\end{figure}

Usually STIRAP is performed in a $\Lambda$-system consisting of two low-energy levels, $|0\rangle$ and $|1\rangle$, and one at high-energy, $|2\rangle$. The energy of $|0\rangle$ is set to zero and those of $|1\rangle$ and $|2\rangle$ are $\Delta$ and $\epsilon$, respectively. Ideally the energy gap $\Delta$ is small compared to $\epsilon$.  In the STIRAP process two coupling-pulses, 
modulated with  amplitudes $\Omega_0(t)$ and $\Omega_1(t)$, are applied
 in resonance with the energy difference between level $|2\rangle$ and the two low-lying levels $|0\rangle$ and  $|1\rangle$, respectively.
 In order to transfer population from the state $|0\rangle$ to $|1\rangle$, the two pulses are applied in counter-intuitive order, namely first the pulse $\Omega_1(t)$ and then $\Omega_0(t)$. A sketch of the three-level-system with the coupling-pulses is shown in Figure \ref{fig:TLFensemble}.

 The STIRAP protocol is known to be robust against detuning of the coupling-pulses from the  $|0 \rangle$-$|2\rangle$-transition and the  $|1 \rangle$-$|2\rangle$-transition, as long as there is two-photon-resonance and both coupling-pulses are detuned by the same frequency $\Delta_{tp}$. Here for simplicity we only consider full resonance since the main effect we are interested in, the effect of a two-level-fluctuator is not qualitatively altered in the case of two-photon-resonance as will be explain below.

Below we assume the coupling-pulse amplitudes $\Omega_0(t)$ and $\Omega_1(t)$ 
to have Gaussian profiles. Both pulses are applied on the time scale $T$. The coupling-pulse $\Omega_0(t)$ is applied after the pulse $\Omega_1(t)$ with a delay of $\tau$. The whole STIRAP process is assumed to take place in the time span $0 \le t \le T_{\textrm{max}}$.
Specifically, we will assume for the numerical simulations~\cite{Ivanov2004} 
\begin{eqnarray}
	\label{eqn:pulses}
\Omega_{0/1}(t) & = & \Omega_{\textrm{max}}\, \exp \left\{ -\frac{\left[ \left( t-\frac{T_{\textrm{max}}}{2} \right) \mp \frac{T}{4}\right]^2}{T^2} \right\}  \nonumber \\
\end{eqnarray}
with cut-offs for $t<0$ and $t>T_{\rm max}$, and we choose $T = \frac{\sqrt{2}}{10}T_{\textrm{max}}$. The pulse sequence is shown in Figure \ref{fig:TLFensemble}.

In the interaction picture, after the rotating wave approximation (RWA) the Hamiltonian reduces to
\begin{equation}
	\label{eqn:STIRAPHam}
	H_S(t) = \left( \begin{array}{c c c}
		0 & 0 & \Omega_0(t) \\
		0 & 0 & \Omega_1(t) \\
		\Omega_0(t) & \Omega_1(t) & 0 \end{array} \right) \ .
\end{equation}
Its eigenstates are traditionally called bright states ($|+\rangle$, $|-\rangle$) and  dark state $|ds\rangle$. The latter is a superposition of 
the states $|0\rangle$ and $|1\rangle$,
\begin{equation}
	\label{eqn:darkstate}
	|ds\rangle = \cos \Theta \  |0\rangle + \sin \Theta \ |1\rangle ,
\end{equation}
with angle $\Theta = \arctan\left( \Omega_1(t)/\Omega_0(t) \right)$, which reduces to $\Theta=0$ for $\Omega_0 = 0$ but $\Omega_1 \ne 0$ and to $\Theta=\pi/2$ for $ \Omega_1 = 0$ but $\Omega_0 \ne 0$, i.e., for the counter-intuitive order used in STIRAP the angle $\Theta$ changes from $0$ to $\pi /2$
and, as long as the adiabaticity condition~\cite{MessiahII} 
is met, all population originally prepared in state $|0\rangle$ is transferred to state $|1\rangle$.

The effect of dissipation on the STIRAP process 
has been modelled in the past by a Lindblad form for the master equation~\cite{ManganoSiewert2008, Paspalakis2004a},
\begin{eqnarray}
	\dot{\rho} &=&  -i[H,\rho] + \mathcal{L}(\rho) \\
	\label{eqn:Lindbladoperator}
	\mathcal{L}(\rho) &=& \sum_{\alpha} \Gamma_{\alpha}[  L_{\alpha} \rho L_{\alpha}^{\dagger}-\frac{1}{2} \{ L_{\alpha}^{\dagger} L_{\alpha} , \rho \} ] \ , 
\end{eqnarray}
where  $\left\{ L_{\alpha} \right\}$ is an orthonormal set of operators and the $\Gamma_{\alpha}$ denote the relaxation and dephasing rates~\cite{Lindblad1976}. 
Extensions to account for the effects of driving on these rates were discussed in Ref.~\onlinecite{Ithier2005}.
In addition, solid state systems frequently are coupled to a structured noise environment due to the presence of strongly coupling two-level fluctuators. In general their effect can not be modelled by the Lindblad equation acting on the three-level-system alone. 

In the STIRAP process the adiabatic rotation depends on the phase coherence between the states $|0\rangle$ and $|1\rangle$. Therefore even pure dephasing causes population to decay from the dark state to the bright states, and as a result it is not transferred to the final state. Although a background bath can also lead to direct relaxation processes we consider in this paper a background that only causes dephasing. As long as the destructive effect of background dephasing can be described by a constant rate, it can be reduced by  performing the STIRAP process faster. It becomes negligible if
$ T_{\textrm{max}}  \ll 1/\Gamma_{\alpha} $
for all dephasing rates in the Lindblad equation.
The adiabaticity condition imposes a further constraint between 
the strength of the coupling-pulses $\Omega_{\textrm{max}}$ and the duration of the process~\cite{Ivanov2004},  $ \Omega_{\textrm{max}}\cdot T_{\textrm{max}} \gg 1$. 
Hence the quality of the STIRAP in the presence of background dephasing effects can be improved by reducing the process time $T_{\textrm{max}}$ and simultaneously increasing $\Omega_{\textrm{max}}$.

The connection between the parameters of Gaussian coupling-pulses and the success rate of STIRAP in the case of pure dephasing has been studied by Ivanov {\sl et al.}~\cite{Ivanov2004} and  Yatsenko {\sl et al.}~\cite{YatsenkoBergmann2002}.
They showed that, as long as the adiabaticity condition is satisfied and the RWA is valid, the population lost from STIRAP depends exponentially on the length $T$ of the pulses and the timescale is set by the dephasing rates, i.e., the transferred population is maximised by decreasing $T$. They
further found for each given coupling-pulse amplitude a lower bound for $T$~\cite{Ivanov2004}. Below this bound
the adiabaticity condition is no longer fulfilled and the population transfer breaks down. The lower bound decreases with increasing  strength of the driving pulses.

In conclusion, as long as the noise can be described by the Lindblad equation, due to the flexibility of the driving pulses in STIRAP, it is possible to achieve a high-quality STIRAP process even in the presence of dephasing. 
For this reason one should choose the laser pulses as strong as possible while staying in the regime of the RWA. We will see that this is no longer the case for a structured environment.

\section{Structured Bath}

\subsection{Model with TLF}

For solid-sate quantum systems the noise environment frequently is structured due to the presence of two-level fluctuators (TLFs)~\cite{Neeley2008, Sousa2009, Lupa2009, Cole2010, *LisenfeldMuller2010}.
In these cases, the Lindblad form described above is not sufficient. 
To proceed we can treat the TLF explicitly as a part of the larger quantum system to be studied.
Alternatively, we can derive from the appropriate system-bath model the corresponding Bloch-Redfield equation~\cite{Carmichael1999}. This approach has been used to study pure dephasing by Shi {\sl et al.}~\cite{Shi2003}. In what follows we will compare the two approaches.

Environmental TLFs are regularly observed in experiments on low-temperature electrical circuits, although their precise microscopic origin is not yet clear. In the context of superconducting qubits, several models for TLFS have been proposed which all fit the experimental data to a greater or lesser extent. Various models are summarised in Ref.\cite{Cole2010} for the case of a phase-qubit, including bistable lattice defects, charge traps and spin impurities. A possible source of TLFs which could be present in all superconducting qubits are two-level tunneling systems\cite{Martinis2005}, bistable lattice defects in the Josephson junctions of the qubit. In this work we use a general model of TLFs as a quantum two level systems that can couple transversally and longitudinally to the three-level-system of interest.  This model has been compared directly to experimental observations\cite{Cole2010, Cole2010, *LisenfeldMuller2010, Bushev2010} and therefore we need not consider the exact microscopic nature of the TLF.

We consider a quantum three-level system coupled to a two-level fluctuator,
 and both coupled to independent baths
$B1$ and $B2$, respectively. We do not assume a specific form for the baths $B1$ and $B2$ but assume that they are generic background baths that lead to constant relaxation and dephasing rates on the three-level-system and the TLF as is described in more detail below. The Hamiltonian of the three-level-system, the TLF and their coupling in the presence of generic background baths is given by 
\begin{eqnarray}
	\label{eqn:SystemBath}
	H(t) &=&  H_{S}(t) + H_{int}(t) + H_{TLF} + H_{B1} + H_{B2} \nonumber \\
	H_{TLF}   &=& - \frac{\omega_0}{2} \tau_z	\, . 
\end{eqnarray}
The Hamiltonian of the driven three-level system $H_{S}(t)$ has been introduced above, eqn.(\ref{eqn:STIRAPHam}). The TLF is described by $H_{TLF}$ and it is assumed to couple transversally to the $|0\rangle \leftrightarrow |1\rangle$ transition, i.e., in the interaction picture, where the driving terms acquire phase factors $e^{\pm i \Delta t}$, within the rotating wave approximation the coupling term is given by
\begin{eqnarray}
	\label{eqn:couplingOp}
	H_{int} & = & g  \left( \begin{array}{c c c}
		0 & \tau^{-} e^{- i  \Delta t} & 0 \\
					\tau^{+} e^{ i \Delta t} & 0 & 0 \\
					0 & 0 & 0 \end{array} \right) \ . \nonumber \\
\end{eqnarray}
To simplify the notation we define new coupling-operators 
\begin{eqnarray}
	\tilde{\tau}^{\pm}(t) &=& \tau^{\pm} e^{\pm i \Delta t} \ .
\end{eqnarray}

The background baths and couplings to the three-level system and the TLF, 
described by $H_{B1}$ and $H_{B2}$,
cause dissipation which we assume is unstructured.
Accordingly they can be characterized by Lindblad terms with relaxation and dephasing rates. 
In the following, for simplicity we ignore pure dephasing and assume low temperatures, i.e., 
we take only the relaxation from the fluctuator state $|\uparrow \rangle$ to  $|\downarrow \rangle$ with rate $\Gamma_\downarrow$ into account.
Furthermore we assume that the direct coupling between the background-bath $H_{B1}$ and the three-level-system is so weak that it can be ignored. Even if this was not the case, the coupling to $H_{B1}$ would only decrease the effectiveness of STIRAP and not alter the qualitative behaviour that arises from the coupling to the TLF.

The equation of motion of the density matrix $\rho$ thus reduces to a 6-dimensional system, which can be solved numerically. Results will be presented below and compared to other approaches.
The analysis can be extended to account for more than one TLF, however, with growing number 
the numerical analysis quickly becomes intractable. Other approaches, such as the 
Bloch-Redfield approach, provide an alternative.

\subsection{Bloch-Redfield Approach}

In the Bloch-Redfield approach the degrees of freedom of the baths, in the present case 
including those of the TLF,  are traced out of the equation of motion for the density matrix of the system \cite{Carmichael1999}. The characteristics of this bath are contained in the 
spectral function, 
i.e., the Fourier-transformed correlation functions of the coupling-operators,
 \begin{eqnarray}
	 C_{\tilde{\tau}^- \tilde{\tau}^+}(\omega) &=& \int_{-\infty}^{\infty} dt \ e^{i \omega t} \langle \tilde{\tau}^-(t)\tilde{\tau}^+(0) \rangle \ .
\end{eqnarray}
The relevant frequencies are the differences between the time-dependent eigenvalues of $H_S(t)$,
\begin{eqnarray}
	\label{eqn:Lambda(t)}
	\omega_{i j}(t) &=& E_{j}(t) - E_{i}(t) \ .
\end{eqnarray}

As usual the master equation for the time evolution of the matrix elements of the density matrix can be written in the eigenbasis of the system Hamiltonian,
\begin{eqnarray}
	\dot{\rho}_{mn} - i (E_m-E_n) \rho_{mn} &=& \sum_{n' m'} \mathcal{R}_{nmn'm'}\rho_{n'm'}  \ ,
\end{eqnarray}
with relaxation and dephasing described by the tensor  $\mathcal{R}_{nmn'm'}$. However, for the driven system considered here, 
the eigenbasis, i.e., the energies $E_m(t)$ and $ \mathcal{R}_{nmn'm'}(t)$ also depend on time. 

Since the construction of the tensor  $ \mathcal{R}_{nmn'm'}(t)$ at every timestep of a numerical simulation would be very time-consuming we use another form for the equation of motion,
\begin{eqnarray}
	\label{eqn:BlochRedfield:app:BR}
	\dot{\rho} =  -i \left[H_{S}(t) , \rho \right] + \mathcal{L}_{\textrm{back}}(\rho) + \mathcal{L}_{\textrm{TLF}}(\rho)   \, , 
		\end{eqnarray}
		with the Lindblad term $\mathcal{L}_{\textrm{back}}(\rho)$ responsible for the effects of the background bath given in equation eq.(\ref{eqn:Lindbladoperator}) where in the case of pure relaxation considered here the prefactor of only one $L_{\alpha}$ is nonzero, $\Gamma_{\downarrow}$ with $L_{\downarrow} = \tau^{-}$ .

The effect of the TLF is contained in	
\begin{eqnarray}	
		\mathcal{L}_{\textrm{TLF}}(\rho) & = & - \frac{1}{4}  \left[  |1 \rangle \langle 0| \ \alpha^{+}_{10}(t) \ \rho - \alpha^{+}_{10}(t) \ \rho \  |1 \rangle \langle 0|\right. \nonumber \\
	& & \quad \left. + \rho \ \alpha^{-}_{01}(t) \  |0\rangle \langle 1| -  |0\rangle \langle 1|\ \rho \ \alpha^{-}_{01}(t) \right]  \ \nonumber \\ \nonumber \\
	\alpha^{\pm}_{kl}(t) & = &  V(t) \left[ \xi^{kl}(t) \circ C_{\tilde{\tau}^- \tilde{\tau}^+}\left( \pm \omega(t) \right) \right] V^{-1}(t) \nonumber \\ \nonumber \\
	 \xi^{kl}(t) &=& V^{-1}(t)  |l\rangle \langle k|V(t) \ .
	\end{eqnarray}
where $\circ$ denotes  element-wise matrix multiplication ($\left[ A \circ B \right]_{ij} = A_{ij} \cdot B_{ij} $), $ C_{\tilde{\tau}^-\tilde{\tau}^+}\left(\omega(t) \right)$ is a matrix defined by,
\begin{eqnarray}
	\left[ C_{\tilde{\tau}^-\tilde{\tau}^+}\left(\omega(t) \right) \right]_{ij} & = &  C_{\tilde{\tau}^-\tilde{\tau}^+}\left(\omega_{ij}(t) \right) \ ,
\end{eqnarray}
and $V(t)$ diagonalizes the Hamiltonian,
\begin{eqnarray}
	V^{-1}(t) H_{S}(t) V(t) & = & \sqrt{\Omega_0^2(t)+\Omega_1^2(t)} \left( \begin{array}{c c c} 
		-1 & 0 & 0 \\
						0 & 0 & 0 \\
						0 & 0 & 1 \\ \end{array}\right) \nonumber \\  \ .
\end{eqnarray}
If we were to allow two-photon-resonance, all eigenenergies would be shifted by $\Delta_{tp}$ and we would need to make the substitution $\sqrt{ \Omega_0^2(t)+\Omega_1^2(t)} \rightarrow \sqrt{\frac{1}{2}\Delta_{tp}^2+\Omega_{0}^2(t)+\Omega_1^2(t)}$.  Similarly, the form of the coupling operators changes in the eigenbasis of the Hamiltonian. This shows that two-photon-resonance would only change the frequency range of the spectral function which is scanned by the system during the STIRAP process. If TLFs are present in both frequency ranges the results are only changed quantitatively and not qualitatively.
 
The spectral functions of the coupling-operators of a single TLF can be obtained from the quantum regression theorem \cite{Carmichael1999, Lax63}, 
and in the interaction picture takes the Lorentzian shape 
\begin{eqnarray}
	\label{eqn:CorrFunTLFRegression}
C_{\tilde{\tau}^- \tilde{\tau}^+}(\omega) &=&  \frac{8 g^2 \Gamma_{\downarrow}}{(\Gamma_{\downarrow})^2 + 4 (\omega - (\omega_0-\Delta))^2 } \ . 
\end{eqnarray}
The width of the peak in the spectral function is given by the rate $\Gamma_{\downarrow}$. If several  TLFs are present that do not couple to each other, the spectral function of the ensemble of TLFs is given by the sum of the spectral functions of the individual TLFs. Therefore we start by considering only one TLF coupling to the three-level system and discuss the case of several TLFs later.  
As the time scale set by the rate $\Gamma_{\downarrow}$ for a certain TLF becomes equal to the timescale of the coherent time evolution of the system, the Markov approximation can no longer be 
applied and the TLF must be treated as part of the system. Such a non-Markovian behaviour has been studied by Kamleitner et al.\cite{Kamleitner2008} in CTAP systems.

We now compare the time evolution of the system during the STIRAP process obtained from the Bloch-Redfield approach with that obtained from the exact result of a numerical simulation of the six-dimensional system, formed by the three level system and the two-level fluctuator. 

The equation of motion of the density matrix of the six-dimensional problem is given by,
\begin{eqnarray}
	\dot{\rho}_6 & =& -i \left[ H_6,\rho_6 \right] + \Gamma_{\downarrow}[  L_{\downarrow, 6} \  \rho_6 \  L_{\downarrow, 6}^{\dagger}-\frac{1}{2} \{ L_{\downarrow, 6}^{\dagger} \ L_{\downarrow,  6}, \ \rho_6 \} ]  \, . \nonumber \\
\end{eqnarray}
The Hamiltonian $H_6$ contains the three-level system, the TLF and their coupling,
\begin{eqnarray}
	H_6 &=& \left( \begin{array}{c c c c c c}
		-\frac{\omega_0}{2} & 0 & 0 & 0 & 0 & 0 \\
		0 & +\frac{\omega_0}{2} & g & 0 & 0 & 0 \\
		0 & g & -\Delta-\frac{\omega_0}{2} & 0 & 0 & 0 \\
		0 & 0 & 0 & -\Delta+\frac{\omega_0}{2} & 0 & 0 \\
		0 & 0 & 0 & 0 & \epsilon-\frac{\omega_0}{2} & 0 \\
		0 & 0 & 0 & 0 & 0 & \epsilon+\frac{\omega_0}{2} \end{array} \right)  \nonumber \\ 
\end{eqnarray}
The six-dimensional operator in the Lindblad part of the equation of motion is
\begin{eqnarray}
	L_{\downarrow,6} & = & \textrm{I}_3 \otimes \tau^- = \left( \begin{array}{c c c c c c}
		0 & 1 & 0 & 0 & 0 & 0 \\
		0 & 0 & 0 & 0 & 0 & 0 \\
		0 & 0 & 0 & 1 & 0 & 0 \\
		0 & 0 & 0 & 0 & 0 & 0 \\
		0 & 0 & 0 & 0 & 0 & 1 \\
		0 & 0 & 0 & 0 & 0 & 0 \end{array} \right)  \, .
\end{eqnarray}

\section{Results and comparison}

\begin{figure}[th!]
	\vspace{0.5cm}
	\includegraphics[scale=0.4]{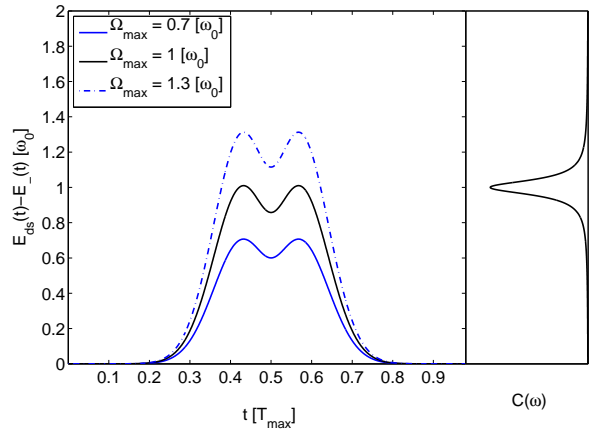}
	\caption{ \label{fig:BohrFrequencies} 
	 The time-dependent difference of the eigenvalues of $H_S(t)$ during STIRAP compared to the spectral function of a TLF with resonance frequency $\omega_0$, shown in the right-hand panel. With increasing maximum pulse strength a resonance is reached.}
\end{figure}

During the STIRAP process the Hamiltonian $H_S(t)$ changes adiabatically,
and so do the system eigenvalues. Consequently, also the strength of decoherence changes,
since it is determined by the spectral function evaluated at the differences of the eigenvalues, 
\begin{eqnarray}
	\omega_{12}(t) &= & \sqrt{\Omega_0^2(t)+\Omega_1^2(t)} \ .
\label{ED}
\end{eqnarray}
The decoherence strength thus depends on the frequency interval covered by the applied pulses.
It becomes strong, when the energy difference eq.(\ref{ED}) is brought into resonance with the TLF's frequency $\omega_0$.
The scenario is illustrated in Figure \ref{fig:BohrFrequencies} for varying strength of  $\Omega_{\textrm{max}}$.

\begin{figure}[t!]
	\vspace{0.5cm}
	\begin{centering}
	\includegraphics[scale=0.4]{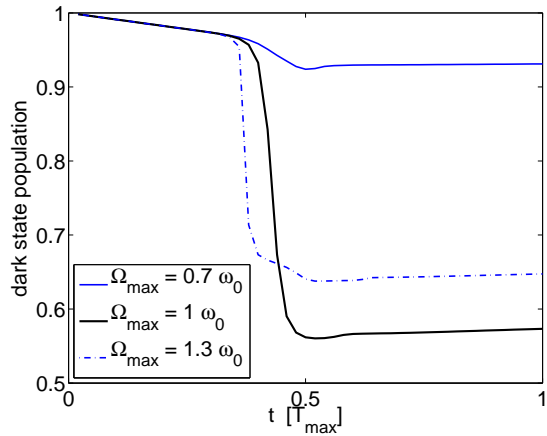}
	\caption{ \label{fig:timeresolvedDSpopulation} 
	The evolution of the population in the dark state during a STIRAP process,
	as obtained from the Bloch-Redfield approach, is plotted for three different pulse amplitudes. 
	For  weak pulses the decoherence is weak during the whole time, and most of the populations stays in the dark state. For $\Omega_{\textrm{max}} = \omega_0$ the system gets in resonance with the TLF during the STIRAP sequence, and the population in the dark state decreases sharply. If the coupling-pulse amplitude is increased further the sharp loss of population starts earlier but lasts for a shorter time, so that a higher fraction of the population remains in the dark state. }
\end{centering}
\end{figure}

If the amplitudes are smaller than $\omega_0$, the TLF remains off-resonance during the whole STIRAP process, and the decoherence it introduces is weak. 
When the maximal difference between the eigenvalues of $H_S(t)$  is resonant with the TLF, the decoherence reaches its highest value. If the coupling-pulse amplitude is increased further, the resonance no longer occurs at the maximal difference of the eigenvalues, 
the system spends less time in resonance with the noise source, and the overall decoherence of the whole process is reduced. The resulting time evolutions of the population of the dark state for three regimes of pulse amplitudes are illustrated in Figure \ref{fig:timeresolvedDSpopulation}.
\begin{figure}[th!]
	\vspace{0.5cm}
	\begin{centering}
	\includegraphics[scale=0.4]{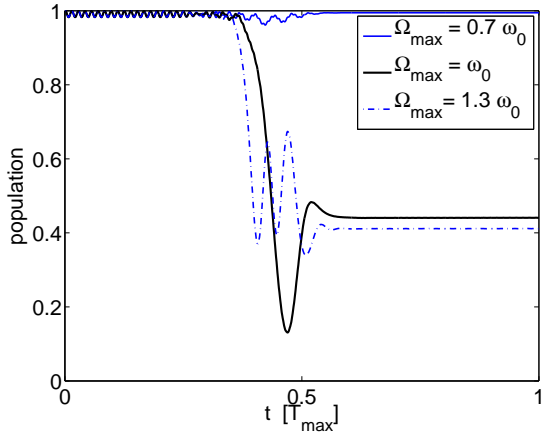}
	\caption{ \label{fig:low-decoherence} The populations of the time-dependent eigenstates, obtained from the six-dimensional simulation,  are plotted for weak damping of the TLF ($\Gamma_{\downarrow} = 0.0007 \omega_0$). At the beginning of the STIRAP sequence when the TLF and the system are off resonance we observe only small oscillations in the population of the states. Once system and TLF are in resonance the population oscillates coherently between the dark and lower lying bright state. The oscillations stop when the system and TLF are again off resonance. }
\end{centering}
\end{figure}

\begin{figure}[t!]
	\vspace{0.5cm}
	\begin{centering}
	\includegraphics[scale=0.4]{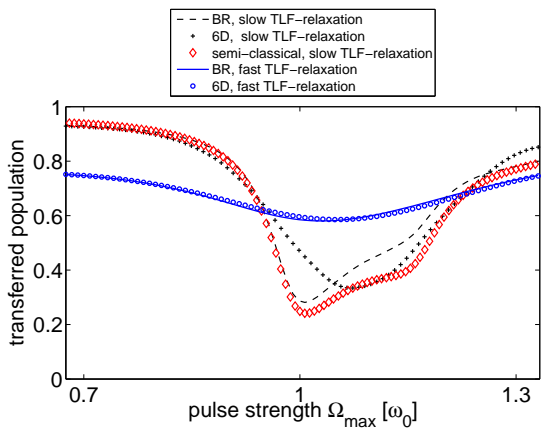}
	\caption{ \label{fig:TransferredPopulation} The transferred population during a STIRAP process is plotted as a function of $\Omega_{\textrm{max}}$. The results of the Bloch-Redfield approach 
	for small $\Gamma_{\downarrow}$ ($\Gamma_{\downarrow} = 0.067 \omega_0$) and large $\Gamma_{\downarrow}$ ($\Gamma_{\downarrow} = 0.33 \omega_0$) are plotted with a dashed line and a solid line, the result of an exact treatment of the TLF  for small $\Gamma_{\downarrow}$ ($\Gamma_{\downarrow} = 0.067 \omega_0$) is drawn with crosses, and the one  for large $\Gamma_{\downarrow}$ ($\Gamma_{\downarrow} = 0.33 \omega_0$) with circles.  The result of the  semi-classical approach for small $\Gamma_{\downarrow}$ ($\Gamma_{\downarrow}= 0.067 \omega_0$) is drawn with a diamond shapes.  }
\end{centering}
\end{figure}

In the case shown in Figure  \ref{fig:low-decoherence}, the coupling between TLF and background bath is so weak that the total system, formed by the three-level-system and the TLF, have a high degree of coherence. In this case the population does not decay from the dark state to the lower lying bright state. Instead we see coherent oscillations between the two states when the system is in resonance with the TLF. As the dephasing rate $\Gamma_{\downarrow}$ is small compared to the inverse timescale of the coherent evolution of the system the TLF has to be treated as part of the system.  The ratio between the population in the two states at the end of the sequence depends predominantly on the coupling strength between system and TLF and the time while system and TLF are in resonance.

Figure \ref{fig:TransferredPopulation} shows how the transferred population decreases 
as one approaches the resonance, and how it recovers as $\Omega_{\textrm{max}}$ is further increased. In this figure we also compare
results obtained from an exact treatment of the TLF and those from the Bloch-Redfield approach.
 They deviate for weak damping, as the effects of the TLF on the system can no longer be described by a Markovian part in the equation of motion containing only the spectral function of the TLF. They coincide for stronger damping when the timescale of relaxation of the TLF is smaller than the timescale of the coherent evolution of the TLF, and the Markov approximation in the derivation of the Bloch-Redfield equation is justified.

 In the strong damping limit, the effect on the transferred population during STIRAP can be described with a semi-classical approximation. We denote the probability of the system to be in the dark state with $p_{ds}(t)$. The transferred population is given by $p_{ds}(T_{\textrm{max}})$. 
We assume that the interaction with the TLF will only lead to relaxation from states with high energy to lower energy states, since  $\Gamma_{\uparrow} = 0$. 
In the adiabatic basis, we obtain a time-dependent relaxation rate $\Gamma_{ds}(t)$ from the dark state to the bright state $|-\rangle$ which yields the differential equation $\dot{p}_{ds}(t) = \Gamma_{ds}(t) p_{ds}(t)$.
The rate $\Gamma_{ds}(t)$ can be approximated by
\begin{eqnarray}
	\Gamma_{ds}(t) = g^2 C_{\tilde{\tau}^- \tilde{\tau}^+}\left[ E_1(t)-E_{2}(t) \right] \ .
\end{eqnarray}
The time evolution of $p_{ds}(t)$ is given by
\begin{eqnarray}
	\label{eqn:rateeqn}
	p_{ds}(t) & =& e^{-A(t)} \nonumber \\
	A(t) & = & g^2 \int_0^t   \frac{\Gamma_{\downarrow}}{\Gamma_{\downarrow}^2+4 \left( \sqrt{\Omega_0^2(t')+\Omega_1^2(t')}-\omega_0 \right)^2} dt' \ . \nonumber \\
\end{eqnarray}
This simplified model reproduces the result of the Bloch-Redfield approach qualitatively as seen in Figure \ref{fig:TransferredPopulation}.

\begin{figure}[t!]
	\vspace{0.2cm}
	\begin{centering}
	\includegraphics[scale=0.4]{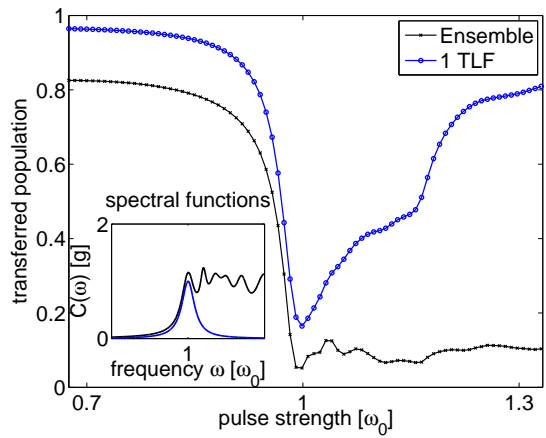}
	\caption{ \label{fig:mergedTLFs} Transferred population as a function of the coupling-pulse strength for two different spectral functions. One Lorentzian spectral function corresponding to a single TLF,
	the second spectral function models the effect of several TLFs. The transferred population recovers from a dip with growing pulse amplitudes if the spectral function decays to zero for higher frequencies, as it is the case for a Lorentzian. In the case of the spectral function that is a superposition of the Lorentzians of several TLFs, little population is transferred once the coupling strength has reached the threshold of the lowest TLF resonance frequency.}
\end{centering}
\end{figure}

The predicted dips in the transferred population could be detected 
by measuring the transferred population in a series of STIRAP experiments with different pulse amplitudes, provided that the dips corresponding to individual TLFs are distinguishable. 
However, for the parameters of TLFs and qubits as observed in experiments \cite{Lupa2009, SembaShnirman2007, LisenfeldMuller2010, Clarke2008} one obtains broad dips. 
In addition, we only obtain sharp dips in a parameter regime where the Bloch-Redfield approach no longer quantitatively reproduces the result of an exact treatment of the two-level fluctuator.
If the system couples to several TLFs with different frequencies the dips in the transferred population merge and  the transferred population just decreases with increasing coupling-pulse strength. 
Figure \ref{fig:mergedTLFs} shows the transferred population for such a dense distribution of TLFs. This limits the possibility of improving the STIRAP process in the presence of background dephasing by reducing the process time and increasing $\Omega_{\textrm{max}}$.

For an efficient implementation of STIRAP  in a system with a small number of TLFs one has to avoid using pulse amplitudes in the vicinity of their frequencies. The intervals to avoid depend on the exact form of the spectral function and the coupling-pulses. 
However, as a rule of thumb one should avoid a  frequency region of two to three times  the width $\Gamma$ of the Lorentzian peak in the spectral function.  

\section{CTAP in the presence of TLFs}
\begin{figure}[t!]
	\includegraphics[scale=0.7]{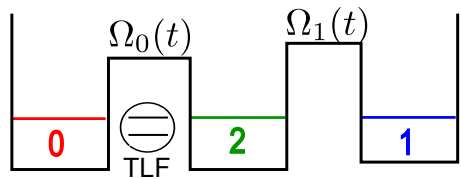}
	\caption{ \label{fig:CTAP-sketch} The three states used in the coherent tunnelling by adiabatic passage (CTAP) correspond to an electron located in one of three neighbouring quantum dots. The TLF is assumed to couple to the time dependent tunneling amplitude which corresponds to the coupling-pulse in STIRAP.}
\end{figure}

In a system consisting of three quantum dots a transfer protocol similar to STIRAP, called coherent tunnelling by adiabatic passage (CTAP), has been proposed\cite{Greentree2004,Kamleitner2008}. The system consists of three potential wells arranged in a line and separated by controllable potential barriers.  Varying the barrier tunneling rates in an analogous way to STIRAP allows an electron to be adiabatically steered from the first to last well without populating the central well. The position of the electron in the three quantum dots corresponds to the three  basis states $|0 \rangle$, $|2\rangle$, and $|1\rangle$ (arranged in this order for similarity with the STIRAP notation). An illustration of the CTAP system is shown in Figure \ref{fig:CTAP-sketch}. The tunneling-matrix-elements between the quantum dots can be tuned by varying the height of the energy barrier between the quantum dots by applying a gate voltage.  We use Gaussian modulations of the tunnelling-matrix-elements which correspond to the microwave coupling-pulses.  CTAP has also been studied in Bose-Einstein condensates\cite{Rab} and has recently been experimentally demonstrated using multi-core optical fibres\cite{Longhi2007}.

In the STIRAP process the microwave coupling-pulses are assumed to be fully coherent.
In CTAP, however, a noisy environment, e.g., TLFs could modify the time-dependent tunnelling-rates between the quantum dots. Hence the pulse amplitudes $\Omega_{0/1}(t)$ controlling the transitions between the dot 0 and 2, or 2 and 1, acquire extra fluctuating terms, $\delta \Omega_{0/1}(t)$. While this 
picture fixes the coupling terms in the basis of the three
dot levels, it could be either longitudinal (diagonal) of transverse (off-diagonal) in the eigenbasis of the TLF. We will see that the two types of coupling cause qualitatively different behaviour
and allow for both, writing
$\delta \Omega_{i}(t)=g_{i}^{x,z} \tau_{x,z}$. If for simplicity we assume that all three states of the quantum dots are energetically degenerate, the noise on the coupling-pulses takes the form 
\begin{eqnarray}
	H_{SB} =  \left( \begin{array}{c c c}
		0 & 0 & g_1^{x,z} \tau_{x,z} \\
		0 & 0 & g_2^{x,z} \tau_{x,z} \\
		g_1^{x,z}\tau_{x,z} & g_2^{x,z} \tau_{x,z} & 0 \end{array} \right) \ .
\end{eqnarray}
For the present example we assume that the relaxation rates $\Gamma_{\downarrow}$ and $\Gamma_{\uparrow}$ of the TLF are equal, which corresponds to a high-temperature environment.

First we consider the case of longitudinal coupling.
With a pure $\tau_z$ coupling we obtain from the quantum regression theorem the spectral function 
\begin{eqnarray}
	C_{\tau_z \tau_z}(\omega) &=& \frac{8 (g_1^{z})^2 \Gamma_{\downarrow}\cdot \Gamma_{\uparrow}}{(\Gamma_{\downarrow}+ \Gamma_{\uparrow})^3+(\Gamma_{\downarrow}+ \Gamma_{\uparrow})\cdot \omega^2} \ .
\end{eqnarray}
It is peaked around zero frequency, and its resonance condition does not depend on the eigenfrequency of the fluctuator. The TLF is in resonance with the difference between the eigenvalues of $H_S(t)$ at the beginning and at the end of the pulse sequence seen in Figure \ref{fig:BohrFrequencies}. Therefore, in this case, the strength of decoherence is not increased as the difference between the eigenvalues of $H_S(t)$ becomes resonant with the TLF. Only the decreasing of the process time $T_{\textrm{max}}$ has an effect on the transferred population as it shortens the time during which the system suffers from strong decoherence.

The transferred population increases with decreasing process time.  One should also note that the efficiency of CTAP is poor for this configuration since the strong decoherence at zero frequency is always present. This is equivalent to the ubiquitous low frequency noise seen in many solid-state systems.
\begin{figure}[t!]
	\vspace{0.2cm}
	\begin{centering}
	\includegraphics[scale=0.4]{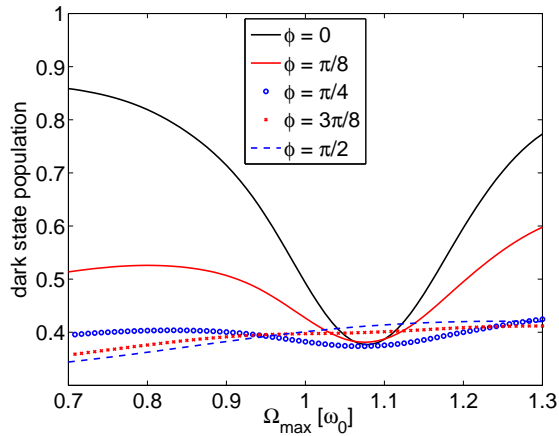}
	\caption{ \label{fig:CTAP-transf-pop} The transferred population is plotted over the coupling-pulse amplitude for a CTAP process where the TLF couples to the coupling-pulse amplitude between states $|0\rangle$ and $|2\rangle$. Different cases of  mixed coupling with the $\sigma_x$-$\tau_z$ coupling amplitude $g \cdot \cos(\phi)$ and  $\sigma_x$-$\tau_x$ coupling amplitude $g \cdot \sin(\phi)$ are plotted. $\Gamma_{\downarrow}$ and $\Gamma_{\uparrow}$ are assumed to be equal. }
\end{centering}
\end{figure}

A qualitatively different behaviour emerges in the presence of transverse coupling.  
With the assumption that there is no relaxation rate from $|\downarrow\rangle$ to $|\uparrow\rangle$ in the TLF ( $\Gamma_{\uparrow} = 0$), the spectral function of the $\tau_x$ operators is the same as that of $\tau^- \tau^+$,
\begin{equation}
	C_{\tau_x,\tau_x}(\omega) = C_{\tau^- \tau^+}(\omega) \ .
\end{equation}

As the spectral function is the same, the behaviour is similar to a TLF that couples transversally to the bare energy levels of the system. We obtain a result that is only quantitatively but not qualitatively different from the behaviour seen in Figure \ref{fig:TransferredPopulation}.
Results of a simulation of a CTAP process with generalised system-TLF coupling,
\begin{eqnarray}
	H_{SB} &=& g \left[ \sin(\phi) \tau_{x} + \cos(\phi) \tau_{z} \right] \left( |0\rangle \langle 2| + |2 \rangle \langle 0 | \right) \ , \nonumber \\
\end{eqnarray}
are shown in Figure \ref{fig:CTAP-transf-pop}. Here we see the smooth transition from the behaviour which is pure dephasing to that which displays TLF resonance effects.

\section{Conclusion}
We analysed the effect of damping on the STIRAP protocol in situations where the environment is dominated by TLFs. In this case the environment is structured and an approach based on a master equation with Lindblad damping terms  acting on the three-level-system alone is not sufficient.
We compared two approaches to account for the structured environment: 
(i) the Bloch-Redfield theory for a three-dimensional system where we account for the effects of the structured environment beyond simple relaxation and dephasing rates by a frequency-dependent spectral function, and (ii) a higher dimensional approach where the TLFs of the environment are treated as part of the system. Generally speaking we can say that the Bloch-Redfield approach works well in the case of a weakly structured environment where the peaks around the TLF frequencies in the spectral function are broad, whereas one has to use the higher dimensional approach for more coherent TLFs.

Due to the time-dependence of the Hamiltonian in STIRAP the system is sensitive to the spectral function of the environment over a wide frequency interval. As a consequence, for a dense distribution of TLFs, the effects of decoherence on the STIRAP process can not be reduced by accelerating the STIRAP process. In the presence of only few TLFs one has to avoid using pulse amplitudes in the vicinity of the TLF resonance frequencies. This problem is more severe for the CTAP protocol, as environmental fluctuators may also couple to the tunnelling-matrix-elements between the quantum dots.

\begin{acknowledgements}
	We want to thank J. Jeske, C. M\"uller, A. Shnirman, and A. Greentree for useful discussions on this work. The work has also been supported in the frame of the EU Information Societies Technologies Programme
``Solid State Systems for Quantum Information Processing (SOLID)''.
\end{acknowledgements}
\bibliography{new}

\begin{thebibliography}{10}%
\makeatletter
\providecommand \@ifxundefined [1]{%
 \ifx #1\undefined \expandafter \@firstoftwo
 \else \expandafter \@secondoftwo
\fi
}%
\providecommand \@ifnum [1]{%
 \ifnum #1\expandafter \@firstoftwo
 \else \expandafter \@secondoftwo
\fi
}%
\providecommand \enquote [1]{``#1''}%
\providecommand \bibnamefont  [1]{#1}%
\providecommand \bibfnamefont [1]{#1}%
\providecommand \citenamefont [1]{#1}%
\providecommand\href[0]{\@sanitize\@href}%
\providecommand\@href[1]{\endgroup\@@startlink{#1}\endgroup\@@href}%
\providecommand\@@href[1]{#1\@@endlink}%
\providecommand \@sanitize [0]{\begingroup\catcode`\&12\catcode`\#12\relax}%
\@ifxundefined \pdfoutput {\@firstoftwo}{%
 \@ifnum{\z@=\pdfoutput}{\@firstoftwo}{\@secondoftwo}%
}{%
 \providecommand\@@startlink[1]{\leavevmode\special{html:<a href="#1">}}%
 \providecommand\@@endlink[0]{\special{html:</a>}}%
}{%
 \providecommand\@@startlink[1]{%
  \leavevmode
  \pdfstartlink
   attr{/Border[0 0 1 ]/H/I/C[0 1 1]}%
   user{/Subtype/Link/A<</Type/Action/S/URI/URI(#1)>>}%
  \relax
 }%
 \providecommand\@@endlink[0]{\pdfendlink}%
}%
\providecommand \url  [0]{\begingroup\@sanitize \@url }%
\providecommand \@url [1]{\endgroup\@href {#1}{\urlprefix}}%
\providecommand \urlprefix [0]{URL }%
\providecommand \Eprint[0]{\href }%
\@ifxundefined \urlstyle {%
  \providecommand \doi [1]{doi:\discretionary{}{}{}#1}%
}{%
  \providecommand \doi [0]{doi:\discretionary{}{}{}\begingroup
  \urlstyle{rm}\Url }%
}%
\providecommand \doibase [0]{http://dx.doi.org/}%
\providecommand \Doi[1]{\href{\doibase#1}}%
\providecommand \bibAnnote [3]{%
  \BibitemShut{#1}%
  \begin{quotation}\noindent
    \textsc{Key:}\ #2\\\textsc{Annotation:}\ #3%
  \end{quotation}%
}%
\providecommand \bibAnnoteFile [2]{%
  \IfFileExists{#2}{\bibAnnote {#1} {#2} {\input{#2}}}{}%
}%
\providecommand \typeout [0]{\immediate \write \m@ne }%
\providecommand \selectlanguage [0]{\@gobble}%
\providecommand \bibinfo [0]{\@secondoftwo}%
\providecommand \bibfield [0]{\@secondoftwo}%
\providecommand \translation [1]{[#1]}%
\providecommand \BibitemOpen[0]{}%
\providecommand \bibitemStop [0]{}%
\providecommand \bibitemNoStop [0]{.\EOS\space}%
\providecommand \EOS [0]{\spacefactor3000\relax}%
\providecommand \BibitemShut [1]{\csname bibitem#1\endcsname}%
\bibitem{Astafiev2007}%
  \BibitemOpen
  \bibfield{author}{%
  \bibinfo {author} {\bibfnamefont{O.}~\bibnamefont{Astafiev}}, \bibinfo
  {author} {\bibfnamefont{K.}~\bibnamefont{Inomata}}, \bibinfo {author}
  {\bibfnamefont{A.~O.}\ \bibnamefont{Niskanen}}, \bibinfo {author}
  {\bibfnamefont{T.}~\bibnamefont{Yamamoto}}, \bibinfo {author}
  {\bibfnamefont{Y.~A.}\ \bibnamefont{Pashkin}}, \bibinfo {author}
  {\bibfnamefont{Y.}~\bibnamefont{Nakamura}},\ and\ \bibinfo {author}
  {\bibfnamefont{J.~S.}\ \bibnamefont{Tsai}},\ }%
  \bibfield{journal}{%
  \bibinfo {journal} {Nature}\ }%
  \textbf{\bibinfo {volume} {449}},\ \bibinfo {pages} {588} (\bibinfo {month}
  {10}\ \bibinfo {year} {2007})%
  \bibAnnoteFile{NoStop}{Astafiev2007}%
\bibitem{Wallraf2008}%
  \BibitemOpen
  \bibfield{author}{%
  \bibinfo {author} {\bibfnamefont{J.~M.}\ \bibnamefont{Fink}}, \bibinfo
  {author} {\bibfnamefont{M.}~\bibnamefont{Goppl}}, \bibinfo {author}
  {\bibfnamefont{M.}~\bibnamefont{Baur}}, \bibinfo {author}
  {\bibfnamefont{R.}~\bibnamefont{Bianchetti}}, \bibinfo {author}
  {\bibfnamefont{P.~J.}\ \bibnamefont{Leek}}, \bibinfo {author}
  {\bibfnamefont{A.}~\bibnamefont{Blais}},\ and\ \bibinfo {author}
  {\bibfnamefont{A.}~\bibnamefont{Wallraff}},\ }%
  \bibfield{journal}{%
  \bibinfo {journal} {Nature}\ }%
  \textbf{\bibinfo {volume} {454}},\ \bibinfo {pages} {315} (\bibinfo {month}
  {07}\ \bibinfo {year} {2008})%
  \bibAnnoteFile{NoStop}{Wallraf2008}%
\bibitem{Martinis2009}%
  \BibitemOpen
  \bibfield{author}{%
  \bibinfo {author} {\bibfnamefont{M.}~\bibnamefont{Hofheinz}}, \bibinfo
  {author} {\bibfnamefont{H.}~\bibnamefont{Wang}}, \bibinfo {author}
  {\bibfnamefont{M.}~\bibnamefont{Ansmann}}, \bibinfo {author}
  {\bibfnamefont{R.~C.}\ \bibnamefont{Bialczak}}, \bibinfo {author}
  {\bibfnamefont{E.}~\bibnamefont{Lucero}}, \bibinfo {author}
  {\bibfnamefont{M.}~\bibnamefont{Neeley}}, \bibinfo {author}
  {\bibfnamefont{A.~D.}\ \bibnamefont{O'Connell}}, \bibinfo {author}
  {\bibfnamefont{D.}~\bibnamefont{Sank}}, \bibinfo {author}
  {\bibfnamefont{J.}~\bibnamefont{Wenner}}, \bibinfo {author}
  {\bibfnamefont{J.~M.}\ \bibnamefont{Martinis}},\ and\ \bibinfo {author}
  {\bibfnamefont{A.~N.}\ \bibnamefont{Cleland}},\ }%
  \bibfield{journal}{%
  \bibinfo {journal} {Nature}\ }%
  \textbf{\bibinfo {volume} {459}},\ \bibinfo {pages} {546} (\bibinfo {month}
  {05}\ \bibinfo {year} {2009})%
  \bibAnnoteFile{NoStop}{Martinis2009}%
\bibitem{Gross2010}%
  \BibitemOpen
  \bibfield{author}{%
  \bibinfo {author} {\bibfnamefont{T.}~\bibnamefont{Niemczyk}}, \bibinfo
  {author} {\bibfnamefont{F.}~\bibnamefont{Deppe}}, \bibinfo {author}
  {\bibfnamefont{H.}~\bibnamefont{Huebl}}, \bibinfo {author}
  {\bibfnamefont{E.~P.}\ \bibnamefont{Menzel}}, \bibinfo {author}
  {\bibfnamefont{F.}~\bibnamefont{Hocke}}, \bibinfo {author}
  {\bibfnamefont{M.~J.}\ \bibnamefont{Schwarz}}, \bibinfo {author}
  {\bibfnamefont{J.~J.}\ \bibnamefont{Garcia-Ripoll}}, \bibinfo {author}
  {\bibfnamefont{D.}~\bibnamefont{Zueco}}, \bibinfo {author}
  {\bibfnamefont{T.}~\bibnamefont{Hummer}}, \bibinfo {author}
  {\bibfnamefont{E.}~\bibnamefont{Solano}}, \bibinfo {author}
  {\bibfnamefont{A.}~\bibnamefont{Marx}},\ and\ \bibinfo {author}
  {\bibfnamefont{R.}~\bibnamefont{Gross}},\ }%
  \bibfield{journal}{%
  \bibinfo {journal} {Nat Phys}\ }%
  \textbf{\bibinfo {volume} {6}},\ \bibinfo {pages} {772} (\bibinfo {month}
  {10}\ \bibinfo {year} {2010})%
  \bibAnnoteFile{NoStop}{Gross2010}%
\bibitem{Marthaler2011}%
  \BibitemOpen
  \bibfield{author}{%
  \bibinfo {author} {\bibfnamefont{M.}~\bibnamefont{Marthaler}}, \bibinfo
  {author} {\bibfnamefont{J.}~\bibnamefont{Lepp\"akangas}},\ and\ \bibinfo
  {author} {\bibfnamefont{J.~H.}\ \bibnamefont{Cole}},\ }%
  \bibfield{journal}{%
  \Doi{10.1103/PhysRevB.83.180505}{\bibinfo {journal} {Phys. Rev. B}}\ }%
  \textbf{\bibinfo {volume} {83}},\ \bibinfo {pages} {180505} (\bibinfo {month}
  {May}\ \bibinfo {year} {2011})%
  \bibAnnoteFile{NoStop}{Marthaler2011}%
\bibitem{Bergmann1998}%
  \BibitemOpen
  \bibfield{author}{%
  \bibinfo {author} {\bibfnamefont{K.}~\bibnamefont{Bergmann}}, \bibinfo
  {author} {\bibfnamefont{H.}~\bibnamefont{Theuer}},\ and\ \bibinfo {author}
  {\bibfnamefont{B.~W.}\ \bibnamefont{Shore}},\ }%
  \bibfield{journal}{%
  \Doi{10.1103/RevModPhys.70.1003}{\bibinfo {journal} {Rev. Mod. Phys.}}\ }%
  \textbf{\bibinfo {volume} {70}},\ \bibinfo {pages} {1003} (\bibinfo {month}
  {Jul}\ \bibinfo {year} {1998})%
  \bibAnnoteFile{NoStop}{Bergmann1998}%
\bibitem{Kis2002}%
  \BibitemOpen
  \bibfield{author}{%
  \bibinfo {author} {\bibfnamefont{Z.}~\bibnamefont{Kis}}\ and\ \bibinfo
  {author} {\bibfnamefont{F.}~\bibnamefont{Renzoni}},\ }%
  \bibfield{journal}{%
  \Doi{10.1103/PhysRevA.65.032318}{\bibinfo {journal} {Phys. Rev. A}}\ }%
  \textbf{\bibinfo {volume} {65}},\ \bibinfo {pages} {032318} (\bibinfo {month}
  {Feb}\ \bibinfo {year} {2002})%
  \bibAnnoteFile{NoStop}{Kis2002}%
\bibitem{ManganoSiewert2008}%
  \BibitemOpen
  \bibfield{author}{%
  \bibinfo {author} {\bibfnamefont{G.}~\bibnamefont{Mangano}}, \bibinfo
  {author} {\bibfnamefont{J.}~\bibnamefont{Siewert}},\ and\ \bibinfo {author}
  {\bibfnamefont{G.}~\bibnamefont{Falci}},\ }%
  \bibfield{journal}{%
  \Doi{10.1140/epjst/e2008-00729-4}{\bibinfo {journal} {Eur. Phys. J. Special
  Topics}}\ }%
  \textbf{\bibinfo {volume} {160}},\ \bibinfo {pages} {259} (\bibinfo {year}
  {2008})%
  \bibAnnoteFile{NoStop}{ManganoSiewert2008}%
\bibitem{Paspalakis2004a}%
  \BibitemOpen
  \bibfield{author}{%
  \bibinfo {author} {\bibfnamefont{K.}~\bibnamefont{Paspalakis},
  \bibfnamefont{E.}}\ and\ \bibinfo {author} {\bibfnamefont{N.~J.}\
  \bibnamefont{Kylstra}},\ }%
  \bibfield{journal}{%
  \Doi{10.1080/09500340408232482}{\bibinfo {journal} {Journal of Modern
  Optics}}\ }%
  \textbf{\bibinfo {volume} {51}},\ \bibinfo {pages} {1679 } (\bibinfo {year}
  {2004})%
  \bibAnnoteFile{NoStop}{Paspalakis2004a}%
\bibitem{Siewert2010}%
  \BibitemOpen
  \bibfield{author}{%
  \bibinfo {author} {\bibfnamefont{J.}~\bibnamefont{Siewert}}, \bibinfo
  {author} {\bibfnamefont{T.}~\bibnamefont{Brandes}},\ and\ \bibinfo {author}
  {\bibfnamefont{G.}~\bibnamefont{Falci}},\ }%
  \bibfield{journal}{%
  \Doi{10.1103/PhysRevB.79.024504}{\bibinfo {journal} {Phys. Rev. B}}\ }%
  \textbf{\bibinfo {volume} {79}},\ \bibinfo {pages} {024504} (\bibinfo {month}
  {Jan}\ \bibinfo {year} {2009})%
  \bibAnnoteFile{NoStop}{Siewert2010}%
\bibitem{Falci2011}%
  \BibitemOpen
  \bibfield{author}{%
  \bibinfo {author} {\bibfnamefont{A.}~\bibnamefont{La~Cognata}}, \bibinfo
  {author} {\bibfnamefont{P.}~\bibnamefont{Caldara}}, \bibinfo {author}
  {\bibfnamefont{D.}~\bibnamefont{Valenti}}, \bibinfo {author}
  {\bibfnamefont{D.}~\bibnamefont{Spagnolo}}, \bibinfo {author}
  {\bibfnamefont{A.}~\bibnamefont{D'Arrigo}}, \bibinfo {author}
  {\bibfnamefont{E.}~\bibnamefont{Paladino}},\ and\ \bibinfo {author}
  {\bibfnamefont{G.}~\bibnamefont{Falci}},\ }%
  \bibfield{journal}{%
  \Doi{10.1142/S0219749911006880}{\bibinfo {journal} {International Journal of
  Quantum Information (IJQI)}}\ }%
  \textbf{\bibinfo {volume} {9}},\ \bibinfo {pages} {1} (\bibinfo {month}
  {Jan}\ \bibinfo {year} {2009})%
  \bibAnnoteFile{NoStop}{Falci2011}%
\bibitem{Hakonen2009}%
  \BibitemOpen
  \bibfield{author}{%
  \bibinfo {author} {\bibfnamefont{M.~A.}\ \bibnamefont{Sillanp\"a\"a}},
  \bibinfo {author} {\bibfnamefont{J.}~\bibnamefont{Li}}, \bibinfo {author}
  {\bibfnamefont{K.}~\bibnamefont{Cicak}}, \bibinfo {author}
  {\bibfnamefont{F.}~\bibnamefont{Altomare}}, \bibinfo {author}
  {\bibfnamefont{J.~I.}\ \bibnamefont{Park}}, \bibinfo {author}
  {\bibfnamefont{R.~W.}\ \bibnamefont{Simmonds}}, \bibinfo {author}
  {\bibfnamefont{G.~S.}\ \bibnamefont{Paraoanu}},\ and\ \bibinfo {author}
  {\bibfnamefont{P.~J.}\ \bibnamefont{Hakonen}},\ }%
  \bibfield{journal}{%
  \Doi{10.1103/PhysRevLett.103.193601}{\bibinfo {journal} {Phys. Rev. Lett.}}\
  }%
  \textbf{\bibinfo {volume} {103}},\ \bibinfo {pages} {193601} (\bibinfo
  {month} {Nov}\ \bibinfo {year} {2009})%
  \bibAnnoteFile{NoStop}{Hakonen2009}%
\bibitem{Astafiev2010}%
  \BibitemOpen
  \bibfield{author}{%
  \bibinfo {author} {\bibfnamefont{O.}~\bibnamefont{Astafiev}}, \bibinfo
  {author} {\bibfnamefont{A.~M.}\ \bibnamefont{Zagoskin}}, \bibinfo {author}
  {\bibfnamefont{A.~A.}\ \bibnamefont{Abdumalikov}}, \bibinfo {author}
  {\bibfnamefont{Y.~A.}\ \bibnamefont{Pashkin}}, \bibinfo {author}
  {\bibfnamefont{T.}~\bibnamefont{Yamamoto}}, \bibinfo {author}
  {\bibfnamefont{K.}~\bibnamefont{Inomata}}, \bibinfo {author}
  {\bibfnamefont{Y.}~\bibnamefont{Nakamura}},\ and\ \bibinfo {author}
  {\bibfnamefont{J.~S.}\ \bibnamefont{Tsai}},\ }%
  \bibfield{journal}{%
  \Doi{10.1126/science.1181918}{\bibinfo {journal} {Science}}\ }%
  \textbf{\bibinfo {volume} {327}},\ \bibinfo {pages} {840} (\bibinfo {year}
  {2010})%
  \bibAnnoteFile{NoStop}{Astafiev2010}%
\bibitem{Wilson2011}%
  \BibitemOpen
  \bibfield{author}{%
  \bibinfo {author} {\bibfnamefont{I.-C.}\ \bibnamefont{Hoi}}, \bibinfo
  {author} {\bibfnamefont{C.~M.}\ \bibnamefont{Wilson}}, \bibinfo {author}
  {\bibfnamefont{G.}~\bibnamefont{Johansson}}, \bibinfo {author}
  {\bibfnamefont{T.}~\bibnamefont{Palomaki}}, \bibinfo {author}
  {\bibfnamefont{B.}~\bibnamefont{Peropadre}},\ and\ \bibinfo {author}
  {\bibfnamefont{P.}~\bibnamefont{Delsing}},\ }%
  \bibfield{journal}{%
  \Doi{10.1103/PhysRevLett.107.073601}{\bibinfo {journal} {Phys. Rev. Lett.}}\
  }%
  \textbf{\bibinfo {volume} {107}},\ \bibinfo {pages} {073601} (\bibinfo
  {month} {Aug}\ \bibinfo {year} {2011})%
  \bibAnnoteFile{NoStop}{Wilson2011}%
\bibitem{Greentree2004}%
  \BibitemOpen
  \bibfield{author}{%
  \bibinfo {author} {\bibfnamefont{A.~D.}\ \bibnamefont{Greentree}}, \bibinfo
  {author} {\bibfnamefont{J.~H.}\ \bibnamefont{Cole}}, \bibinfo {author}
  {\bibfnamefont{A.~R.}\ \bibnamefont{Hamilton}},\ and\ \bibinfo {author}
  {\bibfnamefont{L.~C.~L.}\ \bibnamefont{Hollenberg}},\ }%
  \bibfield{journal}{%
  \Doi{10.1103/PhysRevB.70.235317}{\bibinfo {journal} {Phys. Rev. B}}\ }%
  \textbf{\bibinfo {volume} {70}},\ \bibinfo {pages} {235317} (\bibinfo {month}
  {Dec}\ \bibinfo {year} {2004})%
  \bibAnnoteFile{NoStop}{Greentree2004}%
\bibitem{Clarke2008}%
  \BibitemOpen
  \bibfield{author}{%
  \bibinfo {author} {\bibfnamefont{J.}~\bibnamefont{Clarke}}\ and\ \bibinfo
  {author} {\bibfnamefont{F.~K.}\ \bibnamefont{Wilhelm}},\ }%
  \bibfield{journal}{%
  \bibinfo {journal} {Nature}\ }%
  \textbf{\bibinfo {volume} {453}},\ \bibinfo {pages} {1031} (\bibinfo {month}
  {06}\ \bibinfo {year} {2008})%
  \bibAnnoteFile{NoStop}{Clarke2008}%
\bibitem{Shnirman2005}%
  \BibitemOpen
  \bibfield{author}{%
  \bibinfo {author} {\bibfnamefont{A.}~\bibnamefont{Shnirman}}, \bibinfo
  {author} {\bibfnamefont{G.}~\bibnamefont{Sch\"on}}, \bibinfo {author}
  {\bibfnamefont{I.}~\bibnamefont{Martin}},\ and\ \bibinfo {author}
  {\bibfnamefont{Y.}~\bibnamefont{Makhlin}},\ }%
  \bibfield{journal}{%
  \Doi{10.1103/PhysRevLett.94.127002}{\bibinfo {journal} {Phys. Rev. Lett.}}\
  }%
  \textbf{\bibinfo {volume} {94}},\ \bibinfo {pages} {127002} (\bibinfo {month}
  {Apr}\ \bibinfo {year} {2005})%
  \bibAnnoteFile{NoStop}{Shnirman2005}%
\bibitem{Schriefl2006}%
  \BibitemOpen
  \bibfield{author}{%
  \bibinfo {author} {\bibfnamefont{J.}~\bibnamefont{Schriefl}}, \bibinfo
  {author} {\bibfnamefont{Y.}~\bibnamefont{Makhlin}}, \bibinfo {author}
  {\bibfnamefont{A.}~\bibnamefont{Shnirman}},\ and\ \bibinfo {author}
  {\bibfnamefont{G.}~\bibnamefont{Sch\"on}},\ }%
  \bibfield{journal}{%
  \bibinfo {journal} {New Journal of Physics}\ }%
  \textbf{\bibinfo {volume} {8}},\ \bibinfo {pages} {1} (\bibinfo {year}
  {2006})%
  \bibAnnoteFile{NoStop}{Schriefl2006}%
\bibitem{Martinis2005}%
  \BibitemOpen
  \bibfield{author}{%
  \bibinfo {author} {\bibfnamefont{J.~M.}\ \bibnamefont{Martinis}}, \bibinfo
  {author} {\bibfnamefont{K.~B.}\ \bibnamefont{Cooper}}, \bibinfo {author}
  {\bibfnamefont{R.}~\bibnamefont{McDermott}}, \bibinfo {author}
  {\bibfnamefont{M.}~\bibnamefont{Steffen}}, \bibinfo {author}
  {\bibfnamefont{M.}~\bibnamefont{Ansmann}}, \bibinfo {author}
  {\bibfnamefont{K.~D.}\ \bibnamefont{Osborn}}, \bibinfo {author}
  {\bibfnamefont{K.}~\bibnamefont{Cicak}}, \bibinfo {author}
  {\bibfnamefont{S.}~\bibnamefont{Oh}}, \bibinfo {author}
  {\bibfnamefont{D.~P.}\ \bibnamefont{Pappas}}, \bibinfo {author}
  {\bibfnamefont{R.~W.}\ \bibnamefont{Simmonds}},\ and\ \bibinfo {author}
  {\bibfnamefont{C.~C.}\ \bibnamefont{Yu}},\ }%
  \bibfield{journal}{%
  \Doi{10.1103/PhysRevLett.95.210503}{\bibinfo {journal} {Phys. Rev. Lett.}}\
  }%
  \textbf{\bibinfo {volume} {95}},\ \bibinfo {pages} {210503} (\bibinfo {month}
  {Nov}\ \bibinfo {year} {2005})%
  \bibAnnoteFile{NoStop}{Martinis2005}%
\bibitem{Lupa2009}%
  \BibitemOpen
  \bibfield{author}{%
  \bibinfo {author} {\bibfnamefont{A.}~\bibnamefont{Lupascu}}, \bibinfo
  {author} {\bibfnamefont{P.}~\bibnamefont{Bertet}}, \bibinfo {author}
  {\bibfnamefont{E.~F.~C.}\ \bibnamefont{Driessen}}, \bibinfo {author}
  {\bibfnamefont{C.~J. P.~M.}\ \bibnamefont{Harmans}},\ and\ \bibinfo {author}
  {\bibfnamefont{J.~E.}\ \bibnamefont{Mooij}},\ }%
  \bibfield{journal}{%
  \Doi{10.1103/PhysRevB.80.172506}{\bibinfo {journal} {Phys. Rev. B}}\ }%
  \textbf{\bibinfo {volume} {80}},\ \bibinfo {pages} {172506} (\bibinfo {month}
  {Nov}\ \bibinfo {year} {2009})%
  \bibAnnoteFile{NoStop}{Lupa2009}%
\bibitem{Cole2010}%
  \BibitemOpen
  \bibfield{author}{%
  \bibinfo {author} {\bibfnamefont{J.~H.}\ \bibnamefont{Cole}}, \bibinfo
  {author} {\bibfnamefont{C.}~\bibnamefont{M\"{u}ller}}, \bibinfo {author}
  {\bibfnamefont{P.}~\bibnamefont{Bushev}}, \bibinfo {author}
  {\bibfnamefont{G.~J.}\ \bibnamefont{Grabovskij}}, \bibinfo {author}
  {\bibfnamefont{J.}~\bibnamefont{Lisenfeld}}, \bibinfo {author}
  {\bibfnamefont{A.}~\bibnamefont{Lukashenko}}, \bibinfo {author}
  {\bibfnamefont{A.~V.}\ \bibnamefont{Ustinov}},\ and\ \bibinfo {author}
  {\bibfnamefont{A.}~\bibnamefont{Shnirman}},\ }%
  \bibfield{journal}{%
  \Doi{10.1063/1.3529457}{\bibinfo {journal} {Applied Physics Letters}}\ }%
  \textbf{\bibinfo {volume} {97}},\ \bibinfo {eid} {252501} (\bibinfo {year}
  {2010})%
  \bibAnnoteFile{NoStop}{Cole2010}%
\bibitem{LisenfeldMuller2010}%
  \BibitemOpen
  \bibfield{author}{%
  \bibinfo {author} {\bibfnamefont{J.}~\bibnamefont{Lisenfeld}}, \bibinfo
  {author} {\bibfnamefont{C.}~\bibnamefont{M\"uller}}, \bibinfo {author}
  {\bibfnamefont{J.~H.}\ \bibnamefont{Cole}}, \bibinfo {author}
  {\bibfnamefont{P.}~\bibnamefont{Bushev}}, \bibinfo {author}
  {\bibfnamefont{A.}~\bibnamefont{Lukashenko}}, \bibinfo {author}
  {\bibfnamefont{A.}~\bibnamefont{Shnirman}},\ and\ \bibinfo {author}
  {\bibfnamefont{A.~V.}\ \bibnamefont{Ustinov}},\ }%
  \bibfield{journal}{%
  \Doi{10.1103/PhysRevLett.105.230504}{\bibinfo {journal} {Phys. Rev. Lett.}}\
  }%
  \textbf{\bibinfo {volume} {105}},\ \bibinfo {pages} {230504} (\bibinfo
  {month} {Dec}\ \bibinfo {year} {2010})%
  \bibAnnoteFile{NoStop}{LisenfeldMuller2010}%
\bibitem{Ivanov2004}%
  \BibitemOpen
  \bibfield{author}{%
  \bibinfo {author} {\bibfnamefont{P.~A.}\ \bibnamefont{Ivanov}}, \bibinfo
  {author} {\bibfnamefont{N.~V.}\ \bibnamefont{Vitanov}},\ and\ \bibinfo
  {author} {\bibfnamefont{K.}~\bibnamefont{Bergmann}},\ }%
  \bibfield{journal}{%
  \Doi{10.1103/PhysRevA.70.063409}{\bibinfo {journal} {Phys. Rev. A}}\ }%
  \textbf{\bibinfo {volume} {70}},\ \bibinfo {pages} {063409} (\bibinfo {month}
  {Dec}\ \bibinfo {year} {2004})%
  \bibAnnoteFile{NoStop}{Ivanov2004}%
\bibitem{MessiahII}%
  \BibitemOpen
  \bibfield{author}{%
  \bibinfo {author} {\bibfnamefont{A.}~\bibnamefont{Messiah}},\ }%
  \emph{\bibinfo {title} {Quantum Mechanics Volume II}}\ (\bibinfo {publisher}
  {Elsevier Science B.V.},\ \bibinfo {year} {1961})%
  \bibAnnoteFile{NoStop}{MessiahII}%
\bibitem{Lindblad1976}%
  \BibitemOpen
  \bibfield{author}{%
  \bibinfo {author} {\bibfnamefont{G.}~\bibnamefont{Lindblad}},\ }%
  \bibfield{journal}{%
  \bibinfo {journal} {Commun. Math. Phys.}\ }%
  \textbf{\bibinfo {volume} {48}},\ \bibinfo {pages} {119} (\bibinfo {year}
  {1976})%
  \bibAnnoteFile{NoStop}{Lindblad1976}%
\bibitem{Ithier2005}%
  \BibitemOpen
  \bibfield{author}{%
  \bibinfo {author} {\bibfnamefont{G.}~\bibnamefont{Ithier}}, \bibinfo {author}
  {\bibfnamefont{E.}~\bibnamefont{Collin}}, \bibinfo {author}
  {\bibfnamefont{P.}~\bibnamefont{Joyez}}, \bibinfo {author}
  {\bibfnamefont{P.~J.}\ \bibnamefont{Meeson}}, \bibinfo {author}
  {\bibfnamefont{D.}~\bibnamefont{Vion}}, \bibinfo {author}
  {\bibfnamefont{D.}~\bibnamefont{Esteve}}, \bibinfo {author}
  {\bibfnamefont{F.}~\bibnamefont{Chiarello}}, \bibinfo {author}
  {\bibfnamefont{A.}~\bibnamefont{Shnirman}}, \bibinfo {author}
  {\bibfnamefont{Y.}~\bibnamefont{Makhlin}}, \bibinfo {author}
  {\bibfnamefont{J.}~\bibnamefont{Schriefl}},\ and\ \bibinfo {author}
  {\bibfnamefont{G.}~\bibnamefont{Sch\"on}},\ }%
  \bibfield{journal}{%
  \Doi{10.1103/PhysRevB.72.134519}{\bibinfo {journal} {Phys. Rev. B}}\ }%
  \textbf{\bibinfo {volume} {72}},\ \bibinfo {pages} {134519} (\bibinfo {month}
  {Oct}\ \bibinfo {year} {2005})%
  \bibAnnoteFile{NoStop}{Ithier2005}%
\bibitem{YatsenkoBergmann2002}%
  \BibitemOpen
  \bibfield{author}{%
  \bibinfo {author} {\bibfnamefont{L.~P.}\ \bibnamefont{Yatsenko}}, \bibinfo
  {author} {\bibfnamefont{V.~I.}\ \bibnamefont{Romanenko}}, \bibinfo {author}
  {\bibfnamefont{B.~W.}\ \bibnamefont{Shore}},\ and\ \bibinfo {author}
  {\bibfnamefont{K.}~\bibnamefont{Bergmann}},\ }%
  \bibfield{journal}{%
  \Doi{10.1103/PhysRevA.65.043409}{\bibinfo {journal} {Phys. Rev. A}}\ }%
  \textbf{\bibinfo {volume} {65}},\ \bibinfo {pages} {043409} (\bibinfo {month}
  {Apr}\ \bibinfo {year} {2002})%
  \bibAnnoteFile{NoStop}{YatsenkoBergmann2002}%
\bibitem{Neeley2008}%
  \BibitemOpen
  \bibfield{author}{%
  \bibinfo {author} {\bibfnamefont{M.}~\bibnamefont{Neeley}}, \bibinfo {author}
  {\bibfnamefont{M.}~\bibnamefont{Ansmann}}, \bibinfo {author}
  {\bibfnamefont{R.~C.}\ \bibnamefont{Bialczak}}, \bibinfo {author}
  {\bibfnamefont{M.}~\bibnamefont{Hofheinz}}, \bibinfo {author}
  {\bibfnamefont{N.}~\bibnamefont{Katz}}, \bibinfo {author}
  {\bibfnamefont{E.}~\bibnamefont{Lucero}}, \bibinfo {author}
  {\bibfnamefont{A.}~\bibnamefont{O/'Connell}}, \bibinfo {author}
  {\bibfnamefont{H.}~\bibnamefont{Wang}}, \bibinfo {author}
  {\bibfnamefont{A.~N.}\ \bibnamefont{Cleland}},\ and\ \bibinfo {author}
  {\bibfnamefont{J.~M.}\ \bibnamefont{Martinis}},\ }%
  \bibfield{journal}{%
  \bibinfo {journal} {Nat Phys}\ }%
  \textbf{\bibinfo {volume} {4}},\ \bibinfo {pages} {523} (\bibinfo {month}
  {07}\ \bibinfo {year} {2008})%
  \bibAnnoteFile{NoStop}{Neeley2008}%
\bibitem{Sousa2009}%
  \BibitemOpen
  \bibfield{author}{%
  \bibinfo {author} {\bibfnamefont{R.}~\bibnamefont{de~Sousa}}, \bibinfo
  {author} {\bibfnamefont{K.~B.}\ \bibnamefont{Whaley}}, \bibinfo {author}
  {\bibfnamefont{T.}~\bibnamefont{Hecht}}, \bibinfo {author}
  {\bibfnamefont{J.}~\bibnamefont{von Delft}},\ and\ \bibinfo {author}
  {\bibfnamefont{F.~K.}\ \bibnamefont{Wilhelm}},\ }%
  \bibfield{journal}{%
  \Doi{10.1103/PhysRevB.80.094515}{\bibinfo {journal} {Phys. Rev. B}}\ }%
  \textbf{\bibinfo {volume} {80}},\ \bibinfo {pages} {094515} (\bibinfo {month}
  {Sep}\ \bibinfo {year} {2009})%
  \bibAnnoteFile{NoStop}{Sousa2009}%
\bibitem{Carmichael1999}%
  \BibitemOpen
  \bibfield{author}{%
  \bibinfo {author} {\bibfnamefont{J.}~\bibnamefont{Carmichael},
  \bibfnamefont{Howard}},\ }%
  \emph{\bibinfo {title} {Statistical Methods in Quantum Optics 1}}\ (\bibinfo
  {publisher} {Springer},\ \bibinfo {year} {1999})%
  \bibAnnoteFile{NoStop}{Carmichael1999}%
\bibitem{Shi2003}%
  \BibitemOpen
  \bibfield{author}{%
  \bibinfo {author} {\bibfnamefont{Q.}~\bibnamefont{Shi}}\ and\ \bibinfo
  {author} {\bibfnamefont{E.}~\bibnamefont{Geva}},\ }%
  \bibfield{journal}{%
  \Doi{10.1063/1.1623482}{\bibinfo {journal} {J. Chem. Phys.}}\ }%
  \textbf{\bibinfo {volume} {119}},\ \bibinfo {pages} {11773} (\bibinfo {year}
  {2003}),\ ISSN \bibinfo {issn} {00219606}%
  \bibAnnoteFile{NoStop}{Shi2003}%
\bibitem{Bushev2010}%
  \BibitemOpen
  \bibfield{author}{%
  \bibinfo {author} {\bibfnamefont{P.}~\bibnamefont{Bushev}}, \bibinfo {author}
  {\bibfnamefont{C.}~\bibnamefont{M\"uller}}, \bibinfo {author}
  {\bibfnamefont{J.}~\bibnamefont{Lisenfeld}}, \bibinfo {author}
  {\bibfnamefont{J.~H.}\ \bibnamefont{Cole}}, \bibinfo {author}
  {\bibfnamefont{A.}~\bibnamefont{Lukashenko}}, \bibinfo {author}
  {\bibfnamefont{A.}~\bibnamefont{Shnirman}},\ and\ \bibinfo {author}
  {\bibfnamefont{A.~V.}\ \bibnamefont{Ustinov}},\ }%
  \bibfield{journal}{%
  \Doi{10.1103/PhysRevB.82.134530}{\bibinfo {journal} {Phys. Rev. B}}\ }%
  \textbf{\bibinfo {volume} {82}},\ \bibinfo {pages} {134530} (\bibinfo {month}
  {Oct}\ \bibinfo {year} {2010})%
  \bibAnnoteFile{NoStop}{Bushev2010}%
\bibitem{Lax63}%
  \BibitemOpen
  \bibfield{author}{%
  \bibinfo {author} {\bibfnamefont{M.}~\bibnamefont{Lax}},\ }%
  \bibfield{journal}{%
  \Doi{10.1103/PhysRev.129.2342}{\bibinfo {journal} {Phys. Rev.}}\ }%
  \textbf{\bibinfo {volume} {129}},\ \bibinfo {pages} {2342} (\bibinfo {month}
  {Mar}\ \bibinfo {year} {1963})%
  \bibAnnoteFile{NoStop}{Lax63}%
\bibitem{Kamleitner2008}%
  \BibitemOpen
  \bibfield{author}{%
  \bibinfo {author} {\bibfnamefont{I.}~\bibnamefont{Kamleitner}}, \bibinfo
  {author} {\bibfnamefont{J.}~\bibnamefont{Cresser}},\ and\ \bibinfo {author}
  {\bibfnamefont{J.}~\bibnamefont{Twamley}},\ }%
  \bibfield{journal}{%
  \Doi{10.1103/PhysRevA.77.032331}{\bibinfo {journal} {Phys. Rev. A}}\ }%
  \textbf{\bibinfo {volume} {77}},\ \bibinfo {pages} {032331} (\bibinfo {month}
  {Mar}\ \bibinfo {year} {2008})%
  \bibAnnoteFile{NoStop}{Kamleitner2008}%
\bibitem{SembaShnirman2007}%
  \BibitemOpen
  \bibfield{author}{%
  \bibinfo {author} {\bibfnamefont{K.}~\bibnamefont{Kakuyanagi}}, \bibinfo
  {author} {\bibfnamefont{T.}~\bibnamefont{Meno}}, \bibinfo {author}
  {\bibfnamefont{S.}~\bibnamefont{Saito}}, \bibinfo {author}
  {\bibfnamefont{H.}~\bibnamefont{Nakano}}, \bibinfo {author}
  {\bibfnamefont{K.}~\bibnamefont{Semba}}, \bibinfo {author}
  {\bibfnamefont{H.}~\bibnamefont{Takayanagi}}, \bibinfo {author}
  {\bibfnamefont{F.}~\bibnamefont{Deppe}},\ and\ \bibinfo {author}
  {\bibfnamefont{A.}~\bibnamefont{Shnirman}},\ }%
  \bibfield{journal}{%
  \Doi{10.1103/PhysRevLett.98.047004}{\bibinfo {journal} {Phys. Rev. Lett.}}\
  }%
  \textbf{\bibinfo {volume} {98}},\ \bibinfo {pages} {047004} (\bibinfo {month}
  {Jan}\ \bibinfo {year} {2007})%
  \bibAnnoteFile{NoStop}{SembaShnirman2007}%
\bibitem{Rab}%
  \BibitemOpen
  \bibfield{author}{%
  \bibinfo {author} {\bibfnamefont{M.}~\bibnamefont{Rab}}, \bibinfo {author}
  {\bibfnamefont{J.~H.}\ \bibnamefont{Cole}}, \bibinfo {author}
  {\bibfnamefont{N.~G.}\ \bibnamefont{Parker}}, \bibinfo {author}
  {\bibfnamefont{A.~D.}\ \bibnamefont{Greentree}}, \bibinfo {author}
  {\bibfnamefont{L.~C.~L.}\ \bibnamefont{Hollenberg}},\ and\ \bibinfo {author}
  {\bibfnamefont{A.~M.}\ \bibnamefont{Martin}},\ }%
  \bibfield{journal}{%
  \Doi{10.1103/PhysRevA.77.061602}{\bibinfo {journal} {Phys. Rev. A}}\ }%
  \textbf{\bibinfo {volume} {77}},\ \bibinfo {pages} {061602} (\bibinfo {month}
  {Jun}\ \bibinfo {year} {2008})%
  \bibAnnoteFile{NoStop}{Rab}%
\bibitem{Longhi2007}%
  \BibitemOpen
  \bibfield{author}{%
  \bibinfo {author} {\bibfnamefont{S.}~\bibnamefont{Longhi}}, \bibinfo {author}
  {\bibfnamefont{G.}~\bibnamefont{Della~Valle}}, \bibinfo {author}
  {\bibfnamefont{M.}~\bibnamefont{Ornigotti}},\ and\ \bibinfo {author}
  {\bibfnamefont{P.}~\bibnamefont{Laporta}},\ }%
  \bibfield{journal}{%
  \Doi{10.1103/PhysRevB.76.201101}{\bibinfo {journal} {Phys. Rev. B}}\ }%
  \textbf{\bibinfo {volume} {76}},\ \bibinfo {pages} {201101} (\bibinfo {month}
  {Nov}\ \bibinfo {year} {2007})%
  \bibAnnoteFile{NoStop}{Longhi2007}%
\end{thebibliography}%
\end{document}